\documentclass[prx,aps,superscriptaddress,amssymb,amsmath,twocolumn]{revtex4-1}
\usepackage{amsmath,mathtools}
\usepackage{amssymb}
\usepackage{amsthm}
\usepackage{amsfonts}
\usepackage{listings}
\usepackage{diagbox}
\usepackage{enumerate}
\usepackage{latexsym}
\usepackage{color}
\usepackage{xcolor}
\usepackage{bm}
\usepackage{hyperref}
\usepackage{graphicx}
\usepackage[FIGTOPCAP]{subfigure}
\hypersetup{
 pdfnewwindow=true, colorlinks=true,
 linkcolor=blue, anchorcolor=blue,
 citecolor=blue, filecolor=blue,
 menucolor=blue, urlcolor=blue}

\newcommand{\beginsupplement}{%
        \setcounter{table}{0}
        \renewcommand{\thetable}{S\arabic{table}}%
        \setcounter{figure}{0}
        \renewcommand{\thefigure}{S\arabic{figure}}%
     }

\newcommand{\webirvsp}{\href{https://github.com/zjwang11/irvsp/blob/master/src_irvsp_v2.tar.gz}{\ttfamily Irvsp}}
\newcommand{\webposbr}{\href{https://github.com/zjwang11/UnconvMat/blob/master/src_pos2aBR.tar.gz}{\ttfamily pos2aBR}}
\newcommand{\webspg}{\href{https://spglib.github.io/spglib/}{\ttfamily Spglib}}
\newcommand{\webpho}{\href{https://phonopy.github.io/phonopy/}{\ttfamily phonopy}}

\def\caas{Ca$_2$As}
\def\ele{electride}
\def\eles{electrides}
\def\can{Ca$_2$N}
\def\yc{Y$_2$C}
\def\ll{LaCl}
\def\babi{Ba$_2$Bi}

\def\cap{Ca$_5$P$_3$}

\def\bcn{Ba$_3$CrN$_3$}
\def\nbo{NaBaO}
\def\ko{K$_2$O}
\def\tbp{2b\ (0, 0, $\frac{1}{2})$}

\def\ie{{\it i.e.},\ }

\input{epsf}

\begin{document}
\tolerance 10000

\draft
\title  {Application of topological quantum chemistry in electrides}

\author{Simin Nie}
\thanks{These authors contributed equally to this work.}
\affiliation{Beijing National Laboratory for Condensed Matter Physics,
and Institute of Physics, Chinese Academy of Sciences, Beijing 100190, China}
\affiliation{Department of Materials Science and Engineering, Stanford University, Stanford, California 94305, USA}

\author{Yuting Qian}
\thanks{These authors contributed equally to this work.}
\affiliation{Beijing National Laboratory for Condensed Matter Physics,
and Institute of Physics, Chinese Academy of Sciences, Beijing 100190, China}
\affiliation{University of Chinese Academy of Sciences, Beijing 100049, China}

\author{Jiacheng Gao}
\affiliation{Beijing National Laboratory for Condensed Matter Physics,
and Institute of Physics, Chinese Academy of Sciences, Beijing 100190, China}
\affiliation{University of Chinese Academy of Sciences, Beijing 100049, China}

\author{Zhong Fang}
\affiliation{Beijing National Laboratory for Condensed Matter Physics,
and Institute of Physics, Chinese Academy of Sciences, Beijing 100190, China}
\affiliation{University of Chinese Academy of Sciences, Beijing 100049, China}

\author{Hongming Weng}
\email{hmweng@iphy.ac.cn}
\affiliation{Beijing National Laboratory for Condensed Matter Physics,
and Institute of Physics, Chinese Academy of Sciences, Beijing 100190, China}
\affiliation{University of Chinese Academy of Sciences, Beijing 100049, China}
\affiliation{Songshan Lake Materials Laboratory, Dongguan, Guangdong 523808, China}

\author{Zhijun Wang}
\email{wzj@iphy.ac.cn}
\affiliation{Beijing National Laboratory for Condensed Matter Physics,
and Institute of Physics, Chinese Academy of Sciences, Beijing 100190, China}
\affiliation{University of Chinese Academy of Sciences, Beijing 100049, China}

\begin{abstract}
The recently developed theory of topological quantum chemistry (TQC) has built a close connection between band representations in momentum space and orbital characters in real space. It provides an effective way to diagnose topological materials, leading to the discovery of lots of topological materials after the screening of all known nonmagnetic compounds. On the other hand, it can also efficiently reveal spacial orbital characters, including average charge centers and site-symmetry characters. By using TQC theory with the computed irreducible representations in the first-principles calculations, 
we demonstrate that the \eles~with excess electrons serving as anions at vacancies can be well identified by analyzing band representations (BRs), which \emph{cannot} be expressed as a sum of \emph{atomic-orbital-induced} band representations (aBRs). In fact, the floating bands (formed by the excess electrons) belong to the BRs induced from the ``pseudo-orbitals'' centered at vacancies. In other words, the electrides are proved to be  \emph{unconventional} 
ionic crystals, where a set of occupied bands is not a sum of aBRs but necessarily contains a BR from vacancies. The TQC theory provides a promising avenue to pursue more \ele~candidates in ionic crystals.
\end{abstract}

\maketitle

\section{introduction}
Recently, the discovery of topological materials~\cite{Tang2018Towards,tang2019efficient,zhang2019catalogue,Vergniory2019,po2017symmetry,song2018quantitative,tqc2017,hsieh2012topological,wang2016hourglass,ma2017experimental} has sprung up, and numerous nonmagnetic materials are predicted to be topologically nontrivial by first-principles calculations based on symmetry-based strategies, such as symmetry indicators~\cite{po2017symmetry,song2018quantitative} and topological quantum chemistry (TQC)~\cite{tqc2017}. To be specific, the TQC theory first builds up the character tables for all $k$-points, and compatibility relations for all 230 space groups released on the Bilbao Crystalline Server~\cite{BCSserver}, which make it possible to obtain the corresponding irreducible representations (irreps) of electronic states in the first-principles calculations (\ie \webirvsp~\cite{gao2020irvsp}). 
For a given space group, a certain orbital ($\rho$; labelled by the site-symmetry group) at sites ($q$) can form a set of energy bands in momentum space (labelled by the irreps of $k$-points' little groups). 
This set of irreps at high-symmetry $k$-points is regraded as a BR of $\rho@q$ in the TQC theory.
If a BR $\rho@q$ is a sum of other BRs, it is not elementary; otherwise, it is an elementary BR (eBR). 
Then, it constructs a complete list of (e)BRs, serving as its basic building blocks. By matching the irreps for a 
set of energy bands in a material with those of BRs, one can tell that these bands belong to a certain (e)BR $\rho@q$. As thus, it makes a close link between the irreps in momentum space and the orbital characters in real space. 
The trivial [\emph{resp.} topological (crystalline)] insulators satisfy compatibility relations and can [\emph{resp.} cannot] be expressed as a sum of eBRs, while the enforced topological semimetals violate the compatibility relations. Here, we emphasize that it not only can diagnose topology of energy bands in momentum space, but also can reveal orbital characters in real space (\ie the average charge centers and site-symmetry characters) by doing the BR decomposition for a set of energy bands. It is also noted that for topologically trivial insulators, we can further define the \emph{unconventional} ones, whose BR decomposition has to contain an essential BR induced from a ``pseudo-orbital'' centered at a vacancy, which can be characterized by real-space invariants~\cite{song2020rsi}. In other words, the occupied bands of the \emph{unconventional} insulators (also known as obstructed atomic limits~\cite{tqc2017,cano2018building}) cannot be decomposed as a sum of aBRs (defined as the BR induced by the real atomic orbitals in crystals). To diagnose (topologically trivial) \emph{unconventional} materials by doing the BR decomposition can be widely used in materials science~\cite{todo2020,sheng2019two,lee2020two,hirayama2018electrides}, such as hydrogen storage materials, higher-order topological insulators (HOTIs), and \eles.

We here focus on the TQC application in the \eles. Electrides are defined as ionic crystals with excess electrons confined in particular vacancies~\cite{dawes1986first,dye1987physical,matsuishi2003high,lee2013dicalcium}, whose arrangement determines the properties of~\eles~and gives their classification by the dimensionality~\cite{tada2014high,zhang2017computer,tada2017first,wang2017exploration}. 
The anionic electrons, being not attached to any atom, bond or molecule, exhibit high electron mobility and low work function, which have been experimentally confirmed in Ca$_2$N~\cite{lee2013dicalcium}: a two dimensional (2D)~\ele~with an excess electron per unit cell ([Ca$_2$N]$^+ \cdot e^-$). The low work function of \eles~is beneficial to induce band inversion, and to realize nontrivial band topology~\cite{hirayama2018electrides}. 
Recently, the concept of \eles~with nontrivial band topology has attracted much attention for promising applications in quantum devices~\cite{huang2018topological,hirayama2018electrides,nie2020six,zhu2019computational,zhang2019topological,nie2019topological}. Given the close relation between BRs and orbital characters, TQC may shed light on the origin of \eles~in ionic materials.

Although there are many studies on searching for inorganic \eles, most of them are limited to compute charge distributions~\cite{tada2014high,zhang2017computer,tada2017first,wang2017exploration}. The symmetry analysis of electrides is lacking.
In this work, the BR analysis of TQC theory provides a way to identify the origin of the bands around the Fermi level ($E_F$) from their symmetry eigenvalues (or irreps) alone, which can be used to find new~\ele~candidates effectively. We first introduce the concept of aBRs, 
which denote the BRs induced by real atomic orbitals in crystals. Then we demonstrate that the \eles~are unconventional ionic crystals, whose occupied bands are decomposed as a sum of eBRs, but not a sum of aBRs. It is because that the BR decomposition necessarily contains a BR from vacancies (occupied by excess electrons). Lastly, some potential \eles~are predicted in ionic crystals by the BR analysis of TQC theory, which can be further checked in future experiments.

\section{CALCULATION METHOD}

The Vienna {\it ab initio} simulation package (VASP)~\cite{vasp1,vasp2} with the projector augmented wave method~\cite{paw1,paw2} based on density functional theory was employed for the first-principles calculations. The generalized gradient approximation of Perdew-Burke-Ernzerhof type~\cite{pbe} was used for exchange-correlation potential. The cut-off energy for plane wave expansion was set to 500 eV. A $7\times 7\times 7$ Monkhorst-Pack grid for Brillouin zone sampling was supplied in the self-consistent process. A width of 0.02 eV was adopted in the Gaussian smearing method for Fermi level determination. 
All the band structures were calculated without considering spin-orbit coupling.
In order to get more reliable band structures for the Ca$_2$As family, more accurate calculations with modified Becke-Johnson (mBJ) potential~\cite{tran2009accurate} were performed. The virtual crystal approximation~\cite{nordheim1931electron} was employed to study the electronic structures of K$_2$O$_{0.5}$F$_{0.5}$ and Sr$_6$Ga$_{0.5}$Ge$_{0.5}$N$_5$. The irreps are 
computed by the program \webirvsp~\cite{gao2020irvsp,link1}, and the list of aBRs is given by the program \webposbr~\cite{link2} (see details in Supplementary Material \cite{supp}).

\section{RESULTS AND DISCUSSION}
To achieve an \ele, it is empirically known that three criteria should be satisfied: 
excess electrons, lattice vacancies and suitable electronegativity of the elements. The previous search~\cite{tada2014high,zhang2017computer,tada2017first,wang2017exploration} for \eles~have been done mainly by analyzing the charge density around $E_F$, where electron localization function analysis has proved to be effective. Here, by simply analyzing their symmetry eigenvalues (or irreps) alone at several maximal high-symmetry $k$-points in first-principles calculations, the BR analysis of TQC theory leads to the well understanding of three characteristics of \eles. First, the floating bands are induced from the BRs of vacancies, indicating that their average charge densities are located at the vacancies in real space. Second, due to the loose confinement, the floating bands are usually close to the $E_F$, which is beneficial to induce the band inversion and nontrivial band topology. Third, the excess anionic electrons in vacancies present a strong hydrogen affinity. The absorption of hydrogen usually moves those floating bands far below $E_F$ and lowers the total energy.

\begin{figure}[!b]
\centering
\includegraphics[width=8 cm]{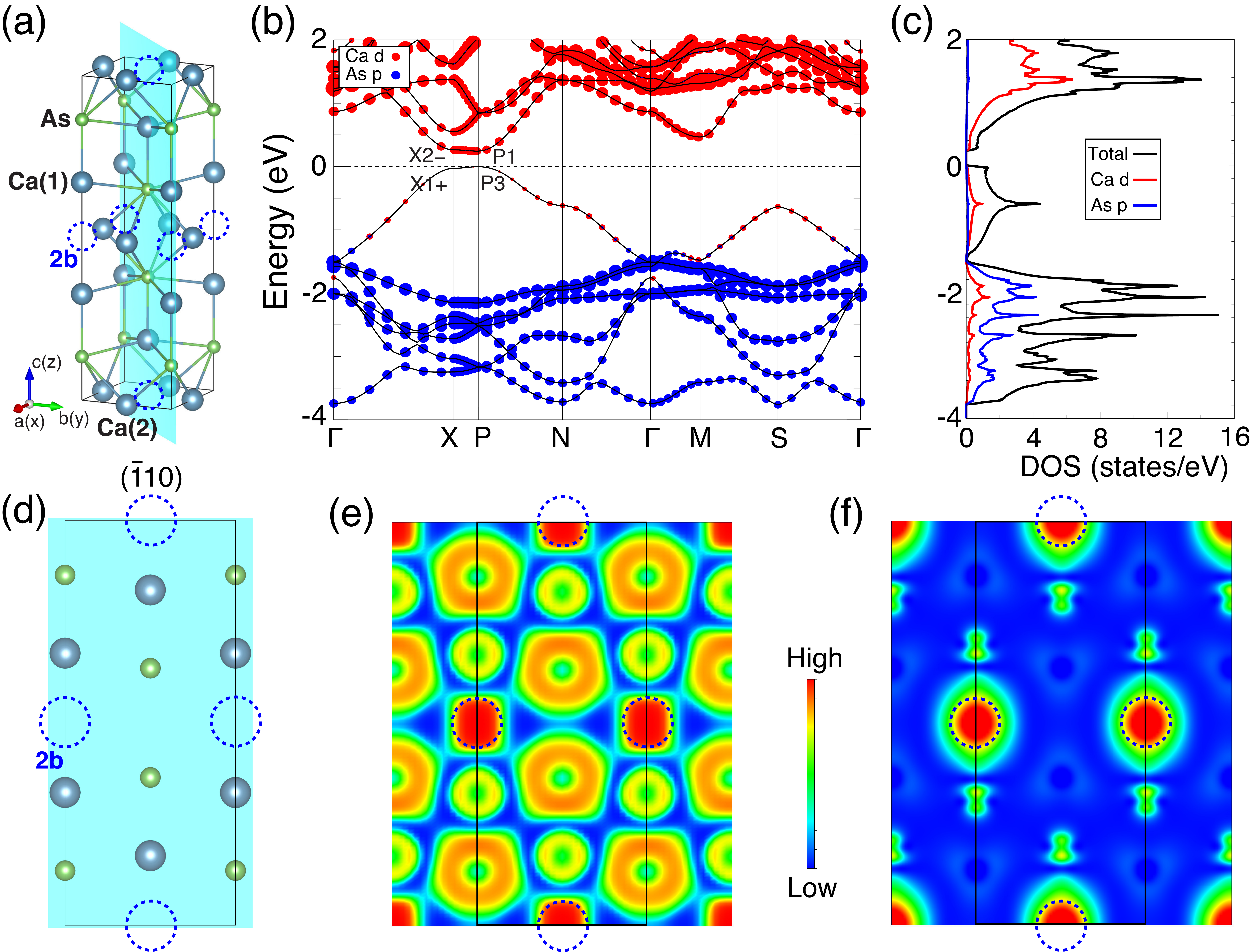}
\caption{(Color online)
(a) Crystal structure and (d) ($\bar110$) lattice plane of Ca$_2$As. (b) The mBJ band structure of~\caas~with spectral weights of Ca-$d$ and As-$p$ orbitals represented by the size of the red and blue circles, respectively. (c) The orbital-resolved DOS of~\caas. (e) Electron localization function of the total electron density of Ca$_2$As, and (f) PED for the states in the energy range -1.3 eV to 0 eV on the ($\bar110$) plane of Ca$_2$As. Hereafter, the corresponding vacancy sites (\ie WKS $2b$ for \caas) are depicted by  blue-dashed circle.
} \label{fig:1}
\end{figure}

\subsection{Band representation of the pseudo-orbitals at voids}
We are interested in the compounds in the \caas~family. The crystal \caas~has a body-centered tetragonal structure and space group $I4/mmm$ (\#139). The lattice parameters are $a=4.63$~\AA, $c= 15.56$~\AA. The As and Ca(1) occupy the Wyckoff sites (WKS) $4e$ (0, 0, $z$) with $z= 0.135$ and $0.328$, respectively, while Ca(2) is at the $4c$ (0, $\frac{1}{2}$, 0), as shown in Fig.~\ref{fig:1} and Table~\ref{tablem:1}. It is worth noting that the~\tbp~positions denoted by blue-dashed circles are hollow. In the $(001)$ planes, the Ca and As atoms form 2D square lattices of Ca and As, respectively, which are stacked alternatively along the $z$-direction, as shown in Fig.~\ref{fig:1}(a).

\begin{table}[!t]
\caption{
Atomic positions and aBRs of the compound Ca$_2$As.
}\label{tablem:1}
\begin{tabular}{c|c|c|c|cc|c}
\hline
\hline
 Atom &WKS($q$) &  Site symm.& Conf. &\multicolumn{2}{c|}{Irreps($\rho$)}& aBRs($\rho@q$)\\
\hline
As&$4e$& $4mm$  & $4p^3$ & $p_z$&$:A_1$   &$A_1@4e$\\
    &        &  &      &$p_x,p_y$&$:E$& $E@4e$\\
  \hline
Ca(1)& $4e$& $4mm$  &$4s^2$&$s$&$:A_1$    & $A_1@4e$\\
  \hline
Ca(2)& $4c$& $mmm$  &$4s^2$ & $s$&$:A_{g}$   & $A_{g}@4c$\\
\cline{5-7}   
\hline
\hline
\end{tabular}
\end{table}

\begin{table}[!t]
\caption{
Irreps and BRs for the 7 valence bands of Ca$_2$As. The BR colored in blue is generated by an excess electron e$^-$ located at vacancies. Hereafter, the irreps are given in the order of increasing energy eigenvalues, and the number in the bracket denotes the degeneracy of the irrep.
}\label{tablem:2}
\begin{tabular}{c|c|c|c|c|c}
\hline
\hline
 & $\Gamma$ (GM) & M & P &X &N \\
\hline
      & GM1+(1)& M1+(1)& P5(2) & X4--(1)& N2--(1)\\
      & GM5+(2)& M3--(1)& P5(2) & X3--(1)& N1+(1)\\
Bands & GM1+(1)& M5+(2)& P1(1) & X4+(1)& N1+(1)\\
      & GM3--(1)& M5--(2)& P3(1) & X1+(1)& N2+(1)\\
      & GM5--(2)& M1+(1)& P3(1) & X3+(1)& N2--(1)\\
      &        &        &      & X2--(1)& N1--(1)\\
      &        &        &      & X1+(1)& N2--(1)\\ 
\hline
\hline
BRs \\
\hline
$A_{1}@4e$ & GM3--(1) & M3--(1) &P3(1) & X2--(1)& N2--(1)\\
           & GM1+(1) & M1+(1) &P1(1)& X1+(1)& N1+(1) \\
\hline
        & GM5--(2)& M5--(2)  &P5(2) &X3--(1)        & N1--(1)                \\
$E@4e$  & GM5+(2)& M5+(2)  &P5(2) &X3+(1)        & N1+(1)                \\
        &        &         &    &        X4--(1)&                N2--(1) \\
        &        &         &   &        X4+(1)&                N2+(1) \\
\hline
{\color{blue}$A_{1g}@2b$} & {\color{blue}GM1+(1)} & {\color{blue}M1+(1)} &{\color{blue}P3(1)}&{\color{blue} X1+(1)}& {\color{blue}N2--(1)} \\
\hline
\hline
\end{tabular}
\end{table}

The band structure of~\caas~with mBJ functional is shown in Fig.~\ref{fig:1}(b), where the size of the red and blue circles represents the weights of Ca-$d$ and As-$p$ orbitals, respectively. In addition to the presence of a global band gap in the band structure, one can also find that the valence bands are dominated by As-$p$ orbitals, while the conduction bands are mainly from Ca-$d$ orbitals. Interestingly, since there are 7 valence bands but only 6 As-$p$ orbitals in a primitive cell, one can notice that the highest valence band in the energy range -1.5 eV to 0 eV is not attributed to either As-$p$ or Ca-$d$ orbitals, which are consistent with the orbital-resolved density of states (DOS), as shown in Fig.~\ref{fig:1}(c). In what follows, we will show that the band does not belong to any aBR.

\begin{figure*}[!t]
\centering
\includegraphics[width=15cm]{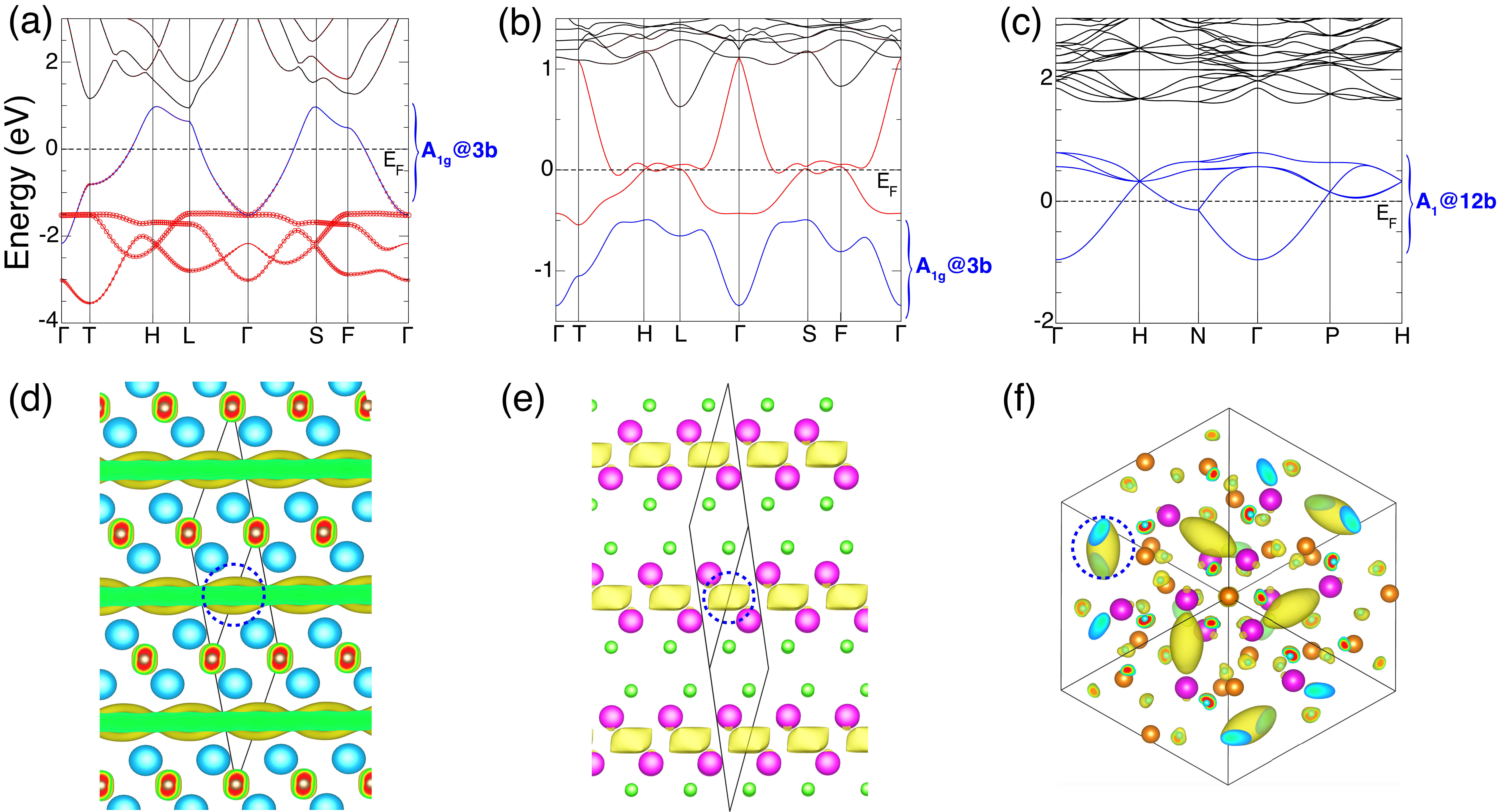}
\caption{(Color online) 
(a-c) The band structures of \can, \ll, and C12A7, respectively. The spectral weight of N $p$ orbitals for~\can~is denoted by the size of the red circles.
(d-f) The  calculated PED of the blue-colored bands for \can, \ll, and C12A7 in the energy ranges of [-1.5 eV, 1 eV], [-1.5 eV, -0.5 eV] and [-1 eV, 1 eV], respectively.
}
\label{fig2}
\end{figure*}

To check the orbital characters of the 7 well-isolated valence bands in the energy range -4 eV to 0 eV, we have computed the irreps for the high-symmetry $k$-point and done the analysis of BRs for these valence bands, which are presented in Table~\ref{tablem:2}. We find that these bands can be decomposed into a sum of BRs: $(A_{1} + E)@4e$ $\oplus$ $A_{1g}@2b$. 
In view of all the aBRs of this compound given in Table~\ref{tablem:1}, the BR analysis indicates that the lowest six bands [\ie $(A_{1} + E)@4e$] are consistent with the states formed by three $p$-orbitals of As (\ie WKS 4e), which agrees well with the fatted band structure [Fig.~\ref{fig:1}(b)]. Interestingly, the remaining band (\ie $A_{1g}@2b$) is contributed from a specific eBR generated by a free electron e$^-$ located at the vacancy $2b$. Therefore, the BR decomposition of \caas~is not a sum of aBRs, suggesting that this inonic compound can be an \ele~candidate in terms of symmetries and irreps. 


The $A_{1g}@2b$ BR of excess electrons is further checked by the calculated charge density distribution. In Figs.~\ref{fig:1}(e) and \ref{fig:1}(f), we plot the electron localization function of all valence states and partial electron density (PED) of the states in the energy range -1.3 eV to 0 eV on ($\bar110$) plane for~\caas, respectively, which both suggest the existence of excess electron distribution at the WKS $2b$. Once the surface termination cuts through the vacancies (\ie the charge centers of the floating bands), surface states would be expected~\cite{hirayama2018electrides,tateishi2020nodal}.
The similar analysis can be also applied to other well-known \eles~\can~\cite{lee2013dicalcium}, \ll~\cite{zhang2017computer}, and Ca$_{24}$Al$_{28}$O$_{64}$ (called C12A7 for short)~\cite{matsuishi2003high} (see details in Sections B and C of the Supplementary Material\cite{supp}). Only using irreps at several high-symmetry $k$-points, the BR analysis of TQC theory tells that the blue bands belong to the BR from vacancies (\ie $A_{1g}@3b$ for Ca$_2$N, $A_{1g}@3b$ for NaCl, and $A_1@2b$ for C12A7 in Fig.~\ref{fig2}), which are formed by excess electrons. 
It is confirmed by projecting the band structure onto the orbital states of the ``empty atom" at the vacancy~\cite{lee2013dicalcium}.


\subsection{Band inversion at the Fermi level}
It is well-known that topological materials usually have the band inversion near $E_F$. Compared with the states constrained by nuclei, the floating states are more conductive, and very close to the $E_F$. Given the very likely presence of the band inversion between the floating bands and other energy bands around $E_F$, it is natural to expect nontrivial band topology in \eles, resulting in the discovery of various topological states in \eles. The irrep and BR analysis will show that the band inversion in \eles~is usually related to the BR of the vacancies. 

\begin{figure}[tb]
\centering
\includegraphics[width=8. cm]{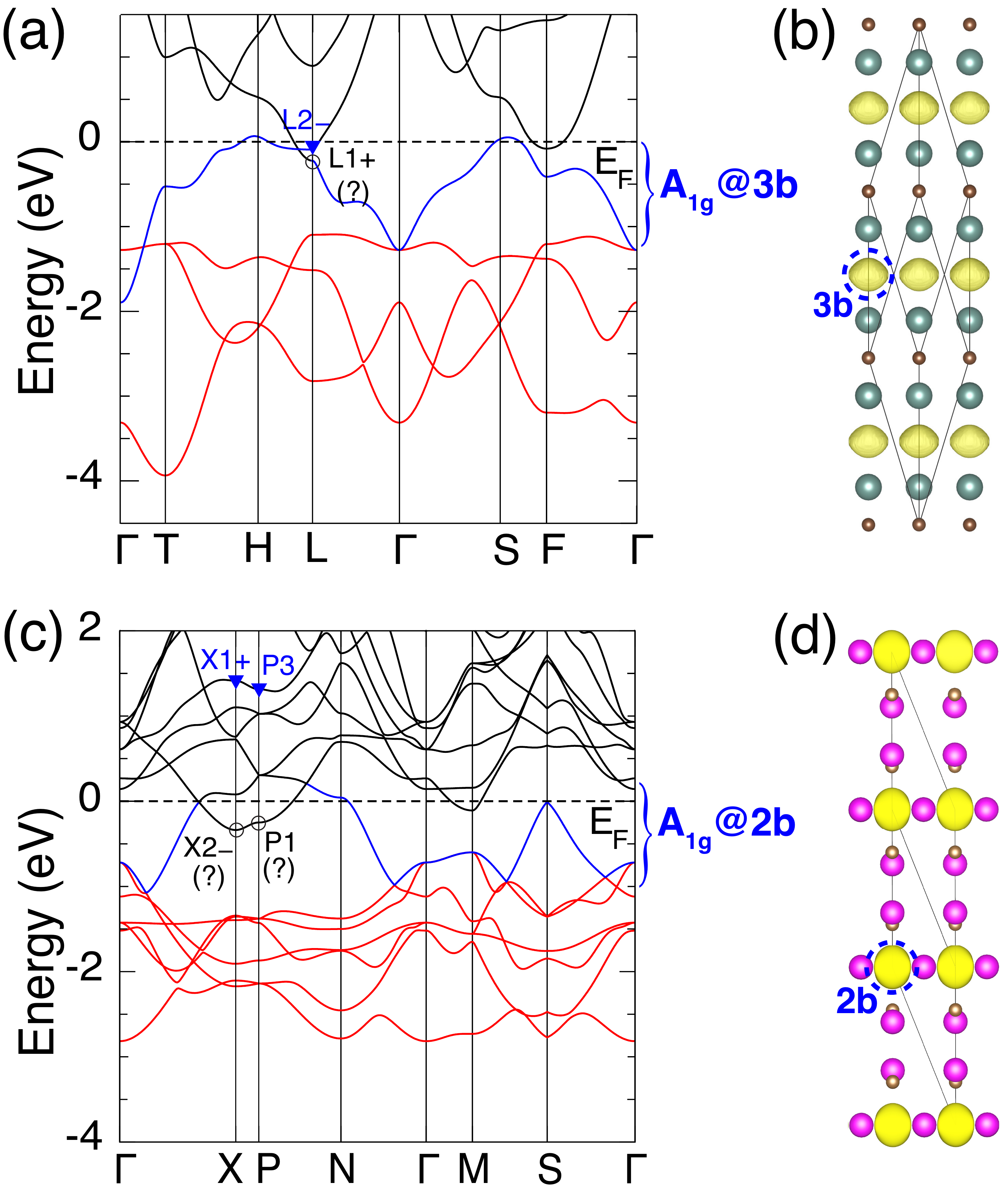}
\caption{(Color online)
(a) Band structures of \yc~with generalized gradient approximation.
(c) Band structures of Ba$_2$Bi with mBJ modification. The irreps L1+, X2--, P1 replaced by ``?'' can be respectively assigned to the irrep L2--, X1+, P3 marked by blue filled triangles after solving the BR decomposition.
The braces show the BRs produced by the bands in the corresponding energy ranges. The PED of the states for \yc~(b) and Ba$_2$Bi (d) in the energy ranges [-1.5 eV, 0 eV] and [-1 eV, 0.5 eV], respectively. 
}
\label{fig:y2c}
\end{figure}

\begin{table}[!b]
\caption{
Atomic positions and aBRs of the compound Y$_2$C.
}\label{table:3}
\begin{tabular}{c|c|c|c|cc|c}
\hline
\hline
Atom &WKS($q$) &  Site symm.& Conf. &\multicolumn{2}{c|}{Irreps($\rho$)}& aBRs($\rho@q$)\\
\hline
Y&$6c$& $3m$  & $5s^25d^2$ &$s$/$d$&:$\rho$& $\rho @ 6c$ \\
  \hline
C & $3a$  & $-3m$ &$2p^2$ & 
$p_z$&$:A_{2u}$ &$A_{2u}$@$3a$   \\
          &    &               &           & $p_x,p_y$&$:E_u$&  $E_u$@$3a$  \\
\hline
\hline
\end{tabular}
\end{table}

Here, we take \yc~with as an example of topological \eles. It crystallizes in
an anti-CdCl$_2$ type structure with the space group of $R\bar3m$ (\#166). 
In the generalized gradient approximation calculation~\cite{pbe} [Fig.~\ref{fig:y2c}(a)], a band inversion is clearly denoted by the irreps (L2-- and L1+) of the two low-energy bands at L, which contributes to a nodal-line structure traversing the full Brillouin zone without spin-orbit coupling~\cite{hirayama2018electrides}.
The aBRs and the analysis of irreps for \yc~are shown in Table \ref{table:3} and Table \ref{table:4}, respectively. 
 After replacing the highest valence band irrep ``L1+'' by ``?''(which denotes an arbitrary irrep), the only solution of BR decomposition for the valence bands in~\yc~is  $(A_{2u} + E_{u}) @3a ~\oplus  A_{1g} @ 3b$ with ``? = L2--''. The aBRs $(A_{2u} + E_{u}) @3a$ originate from C-$p$ orbitals, while the eBR $A_{1g} @ 3b$ is formed by an excess electron at the vacancy site $3b$.
The ``?'' is solved to be ``L2--'', suggesting that the \yc~has a band inversion at L.
The PED of the states colored in blue is also explored to confirm its \ele~feature, as shown in Fig.~\ref{fig:y2c}(b). The presence of charge distribution at the vacancies (\ie WKS $3b$) is consistent with the BR analysis of TQC theory. 

\begin{table}[!t]
\caption{
Irreps and BRs (copied from TQC) for the four valence bands near the Fermi level of Y$_2$C. 
The question mark (``?'') stands for any 1D irrep (\ie the arbitrary irrep) at L. After analyzing the BRs of the occupied bands, one then find that the arbitrary irrep is solved to be L2--(1).
}\label{table:4}
\begin{tabular}{c|c|c|c|c}
\hline
\hline
 & $\Gamma$ & T & F & L  \\
\hline
              &GM2--(1) & T2--(1) &F2--(1)& L2--(1)\\
Bands  &GM1+(1)& T3--(2)&F2--(1)& L2--(1) \\
             & GM3--(2) &T2--(1) &F1--(1)& L1--(1) \\
              &                 &            &F1+(1)&{\color{red} ? (1)}\\
\hline
\hline
BRs \\
\hline
$A_{2u}@3a$& GM2--(1)& T2--(1) &F2--(1)& L2--(1)\\
\hline
$E_{u}@3a$ & GM3--(2)& T3--(2)&F1--(1) $\oplus$ F2--(1)& L1--(1)$\oplus$ L2--(1) \\
\hline
${\color{blue}A_{1g}@3b}$& {\color{blue}GM1+(1)} &{\color{blue}T2--(1)} &{\color{blue}F1+(1)}&{\color{blue} L2--(1)} \\
\hline
\hline
\end{tabular}
\end{table}

The \eles~$A_2B$ ($A=$ Ca, Sr, and Ba; $B=$ As, Sb, and Bi) with the space group of $I4/mmm$ (\#139) are also reported topological nodal-line materials without spin-orbit coupling~\cite{zhang2018hybrid,zhang2019topological}. 
As aforementioned, \caas~is a trivial \ele~with a global band gap.
As the components vary from Ca (As), to Sr (Sb), to Ba (Bi) in the series of materials, the strength of the band inversion changes accordingly.
In~\babi, the band inversion [Fig.~\ref{fig:y2c}(c)] gives rise to a nodal line protected by the coexistence of space inversion and time-reversal symmetries. After replacing the highest valence band irreps X2-- and P1 of Ba$_2$Bi by ``?'', we can obtain the highest valence band from the eBR $A_{1g}@2b$ by solving the BR decomposition. Although the floating band of $A_{1g}@2b$ is not fully occupied and has a band inversion with the lowest conduction band, \babi~can still be considered as an \ele~with nontrivial band topology. As expected, the PED of~\babi~shows the charge densities at the vacancy $2b$ in Fig.~\ref{fig:y2c}(d). In fact, the Ca$_2$As family have lots of compounds, which are all electride candidates (See mBJ band structures for all the compounds in Section D of the Supplementary Material\cite{supp}).

\subsection{Hydrogen absorption}
\begin{figure*}[t]
\centering
\includegraphics[width=12. cm]{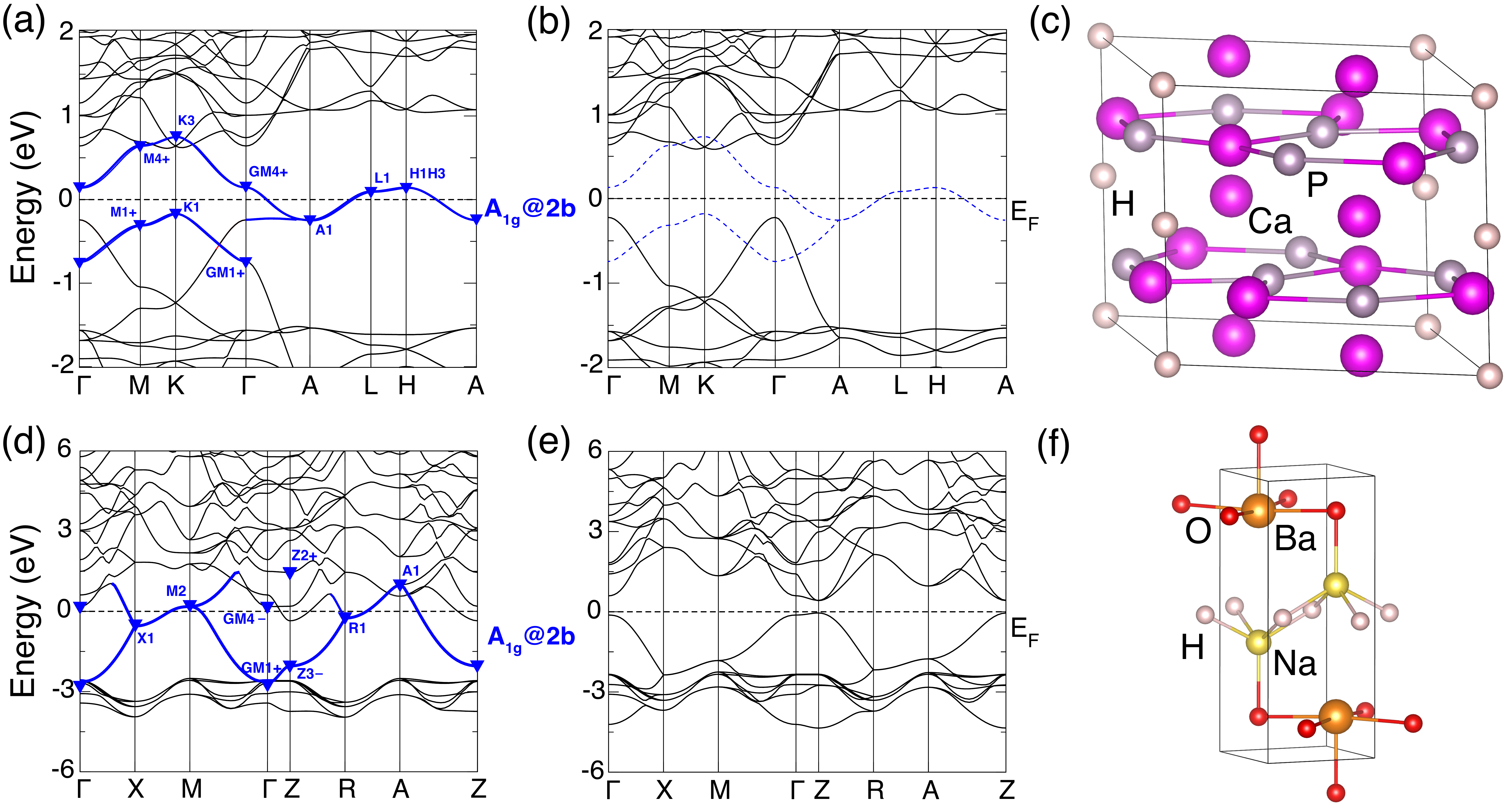}
\caption{(Color online)
The band structures of \cap~(a), \cap H~(b), \nbo~(d) and \nbo H~(e). The floating bands (blue-solid) in~\cap, and \nbo~denote electronic states of excess electrons, whose eBRs can be well determined by the obtained irreps. To facilitate the comparison, the irreps of particular bands are marked out, and the ``disappeared'' floating bands for \cap H are depicted by blue-dashed lines in (b). (c,f) The crystal structures of \cap H and \nbo H, respectively.
}
\label{fig4}
\end{figure*}

The \eles, with excess electrons serving as anions at vacancies, are very easy to absorb hydrogen atoms at the vacancies, which would move the floating bands far away from $E_F$. In order to understand the absorption process, we show the results of \cap~and \bcn~as two paradigms.
\cap~crystallizes in a Mn$_5$Si$_3$-type structure with the space group of P6$_3$/mcm (\#193), and its electronic structure is shown in Fig.~\ref{fig4}(a). The BR analysis shows that these blue-solid bands of~\cap~are the states formed by excess electrons at vacancies, since the eBR $A_{1g} @ 2b$ is determined by the obtained irreps labeled in Fig.~\ref{fig4}(a). After absorption of hydrogen atoms at $2b$ as shown in Fig. \ref{fig4}(c), the floating bands move far below $E_F$ in~\cap H [Fig.~\ref{fig4}(b)], because the hydrogen atoms will bond with the surrounding electrons producing a strong interaction, which is consistent with previous work~\cite{xie2015new}. The similar analysis of Ba$_3$CrN$_3$ can be found in Section~E of the Supplementary Material\cite{supp}. Therefore, the process of adsorbing hydrogen atoms confirms the results of the BR analysis, and strongly supports the presence of the excess electrons at vacancies in \eles.


\subsection{Prediction of \ele~candidates}
Based on the above discussions on three aspects, we find that the BR analysis of TQC theory is very implementable in the understanding of various properties of \eles. 
We demonstrate that the electrides are unconventional ionic crystals, where a set of energy bands is not a sum of aBRs but necessarily contains a BR from vacancies. Guiding by this finding, we apply the BR analysis of TQC to predict some potential \eles~in ionic crystals.

We first propose that NaBaO~\cite{deng2020molecular} with the structure of P4/nmm (\#129) could be an \ele~with an excess electron located at the vacancy $2b$, as the BR of the floating bands [denoted by blue-colored lines and irreps in Fig.~\ref{fig4}(d)] is elementary, and the average charge center is located at the vacancy. In fact, these floating bands are half-filled. By absorbing a hydrogen atom at $2b$ [Fig. \ref{fig4}(f)], the floating bands move completely below $E_F$ [Fig.~\ref{fig4}(e)].

\begin{figure}[!b]
\centering
\includegraphics[width=7.9 cm]{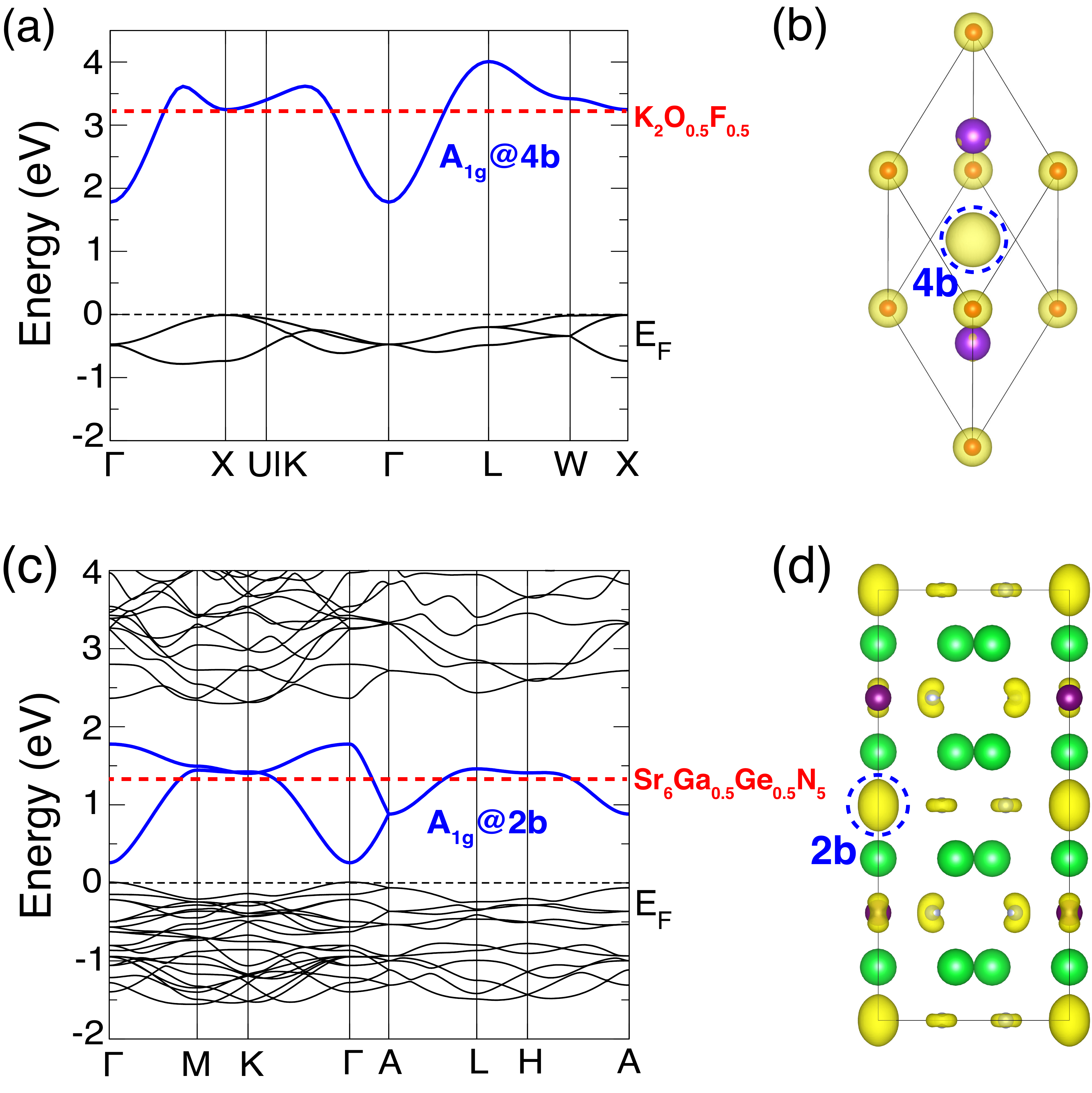}
\caption{(Color online)
The calculated band structures and PED of the floating bands (blue-colored) for K$_2$O (a,b) and Sr$_6$GaN$_5$ (c,d). 
The PED of the states for K$_2$O and Sr$_6$GaN$_5$ in the energy ranges [1 eV, 4 eV] and [0 eV, 2 eV], respectively.
The Fermi levels of K$_2$O$_{0.5}$F$_{0.5}$ and Sr$_6$Ga$_{0.5}$Ge$_{0.5}$N$_5$ are indicated by red-dashed lines, which are obtained by the virtual crystal approximation as implemented in the VASP.}
\label{fig5}
\end{figure}

Next we also find that Sr$_6$GaN$_5$ could be an \ele~after electron doping, \ie K$_2$O$_{0.5}$F$_{0.5}$ and Sr$_6$Ga$_{0.5}$Ge$_{0.5}$N$_5$. 
The band structure of~\ko~with space group $Fm\bar3m$ (\#225)~\cite{tada2014high} is shown in Fig.~\ref{fig5}(a). Above $E_F$, we can see that there is an isolated band (blue-colored) in the energy range 1 eV to 4 eV. With the obtained irreps at high symmetry $k$-points, the BR analysis shows that this band belongs to the eBR of $A_{1g} @ 4b$, which suggests that the band is formed by an excess electron at vacancy $4b$. By replacing 50\% O with F, the isolated band is partially occupied in K$_2$O$_{0.5}$F$_{0.5}$. The presence of a spherical charge distribution around the vacancy in the PED of this band [Fig.~\ref{fig5}(b)] also suggests that K$_2$O$_{0.5}$F$_{0.5}$ is an \ele. 
Heuristically, we propose that Sr$_6$GaN$_5$ with a hexagonal structure of P6$_3$/mcm (\#193) could also be an electride after electron doping. Its band structure is shown in Fig.~\ref{fig5}(c). We find two well-isolated conduction bands (blue-colored) form the eBR of $A_{1g} @ 2b$, indicating the corresponding charge centers at $2b$, which are consistent with the calculated PED [Fig.~\ref{fig5}(d)]. By electron doping, the floating bands of Sr$_6$GaN$_5$ can be reached and become half-filled in 50\% Ge-doped samples (\ie Sr$_6$Ga$_{0.5}$Ge$_{0.5}$N$_5$), as shown in Fig.~\ref{fig5}(c). More details of these two predicted materials, and the results of another candidate (\ie electron-doped Sr$_3$P$_2$) can be found in Section F of the Supplementary Material\cite{supp}. 



\subsection{Discussion}
Through the BR analysis of TQC with the obtained irreps at high-symmetry $k$-points, one can effectively identify the average electronic charge centers and the corresponding site-symmetry characters for a set of separated energy bands, providing an ideal avenue to both understand the essence of the well-known \eles~and find new candidates in the future. Our methodology will facilitate the study of the interesting properties of \eles, such as high electron mobility and low work function. More interestingly, more fundamental physical phenomena can be found in the unconventional materials after considering interactions, such as Sr$_5$P$_3$ and Zr$_5$Sb$_3$.

The compound Sr$_5$P$_3$ has two different structures with the space group of C2/m (\#12; quasi-hexagonal form) and the space group of P6$_3$/mcm (\#193; hexagonal form) at ambient and high pressures, respectively. The atoms in the quasi-hexagonal form deviate somewhat from that of the hexagonal form, and the hexagonal Sr$_5$P$_3$ shares the same crystal structure and similar band structure of Ca$_5$P$_3$. Thus, they are both expected to be \eles~with half-filled floating bands. However, the obtained metallic band structure in the nonmagnetic calculations is in contrast with the observed insulator-type conductivity in experiments~\cite{wang2017exploration}. To remedy this discrepancy, the spin-polarized calculation and Coulomb-U effect have been considered in Ref. \cite{wang2017exploration} and a tiny band gap can be found, resulting in the formation of lower and upper Hubbard bands with DOS peaks below and above E$_F$, respectively. They conjecture that this material may be a Mott insulator due to the effect of the on-site Coulomb interaction. In addition, the phase transition from the hexagonal form (high pressure) to the quasi-hexagonal form (ambient pressure) also suggests the presence of a commensurate charge-density-wave transition.

The compound Zr$_5$Sb$_3$ is experimentally found to be the first superconductor in the large family of compounds with Mn$_5$Si$_3$-type structure (\#193)~\cite{lv2013sup}, which are believed to be due to the electron-phonon coupling. Although the band structure is a little complicated, after carefully checking the irreps of the electronic states~\cite{todo2020}, we find that their irreps could be decomposed into a sum of BRs: $(A+B_1+B_2)@6g$ (Sb atoms), $(A_1+E)@4d$ [Zr(1) atoms] and $A_1{'}@2a$ (which is hollow). By electron counting, the floating bands of eBR $A_1{'}@2a$ are 1/4 filled, and create large Fermi surfaces. 
When the interstitial sites are filled by extra Sb atoms, the experimental measurements have also found that the superconductivity in Zr$_5$Sb$_{3+x}$ is suppressed. Similar results are also confirmed in Zr$_5$Sb$_3$O and Zr$_5$Sb$_{3}$C. These experiments suggest that the superconductivity may be related to the floating electronic states.

\section{CONCLUSIONS}

In conclusion, we demonstrate that the analysis of irreps and BRs in the TQC theory provides an effective way to identify the origin of the energy bands around E$_F$ from their symmetry eigenvalues (or irreps) alone, which is extremely useful to find new~\ele~candidates.
The electrides are proved to be unconventional ionic crystals, where a set of well-separated energy bands below/around the $E_F$ is not a sum of aBRs but necessarily contains a BR from vacancies.
In addition, three characteristics of electrides can be well understood in the TQC theory. First, since there are floating bands with charge densities centered at vacancies in real space, 
surface states could emerge when the surface termination cuts through these vacancies. Second, as the excess electrons show the lack of strong confinement, low work function and floating bands close to $E_F$ are expected. Consequently, the band inversion and nontrivial band topology are very likely to happen in \eles. Third, the interstitial anionic electrons are easy to absorb hydrogen atoms, which would move the floating bands far away from $E_F$ and gain the benefit in total energy. More generally, the hydrogen storage materials and 2D HOTIs can be known as unconventional metal alloys and covalent compounds with trivial occupied bands being not a sum of aBRs, respectively. Anyway, the disagreement between the average charge centers and the atomic positions can be well diagnosed by the BR analysis of TQC theory, which can be widely used in these \emph{unconventional} materials~\cite{todo2020}, such as \eles, hydrogen storage materials and HOTIs.

\noindent \textbf{Acknowledgments} ---
This work was supported by the National Natural Science Foundation of China (Grants No. 11974395), the Strategic Priority Research Program of Chinese Academy of Sciences (Grant No. XDB33000000), and the Center for Materials Genome.
H.W. acknowledges support from the Ministry of Science and Technology of China under grant numbers 2016YFA0300600 and 2018YFA0305700, the Chinese Academy of Sciences under grant number XDB28000000, the Science Challenge Project (No.  TZ2016004), the K. C. Wong Education Foundation (GJTD-2018-01), Beijing Municipal Science \& Technology Commission (Z181100004218001) and Beijing Natural Science Foundation (Z180008).


\bibliography{Ref}

\clearpage
\begin{widetext}
\beginsupplement{}
\setcounter{section}{0}
\section*{SUPPLEMENTARY INFORMATION}

\subsection{The code \webposbr}
\label{sup:a}

The program \webposbr~is developed to get the aBRs from a given crystal structure (\ie POSCAR).
The \webspg~library~\cite{togo2018textttspglib} is needed for the program.
After installing the \webspg~library and copying {\ttfamily libsymspg.a} to the {\ttfamily src\_pos2aBR} folder, the executable binary \webposbr~can be compiled by typing the following commands:
\lstset{language=bash, basicstyle=\ttfamily, frame=shadowbox}
\begin{lstlisting}
   $ ./configure.sh 
   $ source ~/.bashrc
   $ make
\end{lstlisting}

Before running \webposbr, you would better standardize the POSCAR by using \webpho. A typical workflow of \webposbr~is shown as follows:
\lstset{language=bash, keywordstyle=\color{blue!70}, basicstyle=\ttfamily, frame=shadowbox}
\begin{lstlisting}
   $ phonopy --symmetry --tolerance 0.01 -c POSCAR
   $ cp PPOSCAR POSCAR 
   $ pos2aBR
\end{lstlisting}

The program \webposbr~converts ``POSCAR" to ``POSCAR\_std". One should use "POSCAR\_std" to do the calculations within VASP, to be compatible with the program \webirvsp~and TQC work~\cite{bradlyn2017topological,cano2018building,vergniory2017graph,cano2018topology}. In the meantime, its standard output contains the WKS ($q$) for each atom and the irreps ($\rho$) for atomic orbitals (\ie $s$, $p$, $d$). These BRs ($\rho@q$) are the aBRs for the crystal.
An example of \webposbr~output for Y$_2$C is shown in Fig.~\ref{fig:Y2C}.
\begin{figure}[!htb]
  \includegraphics[scale=0.43]{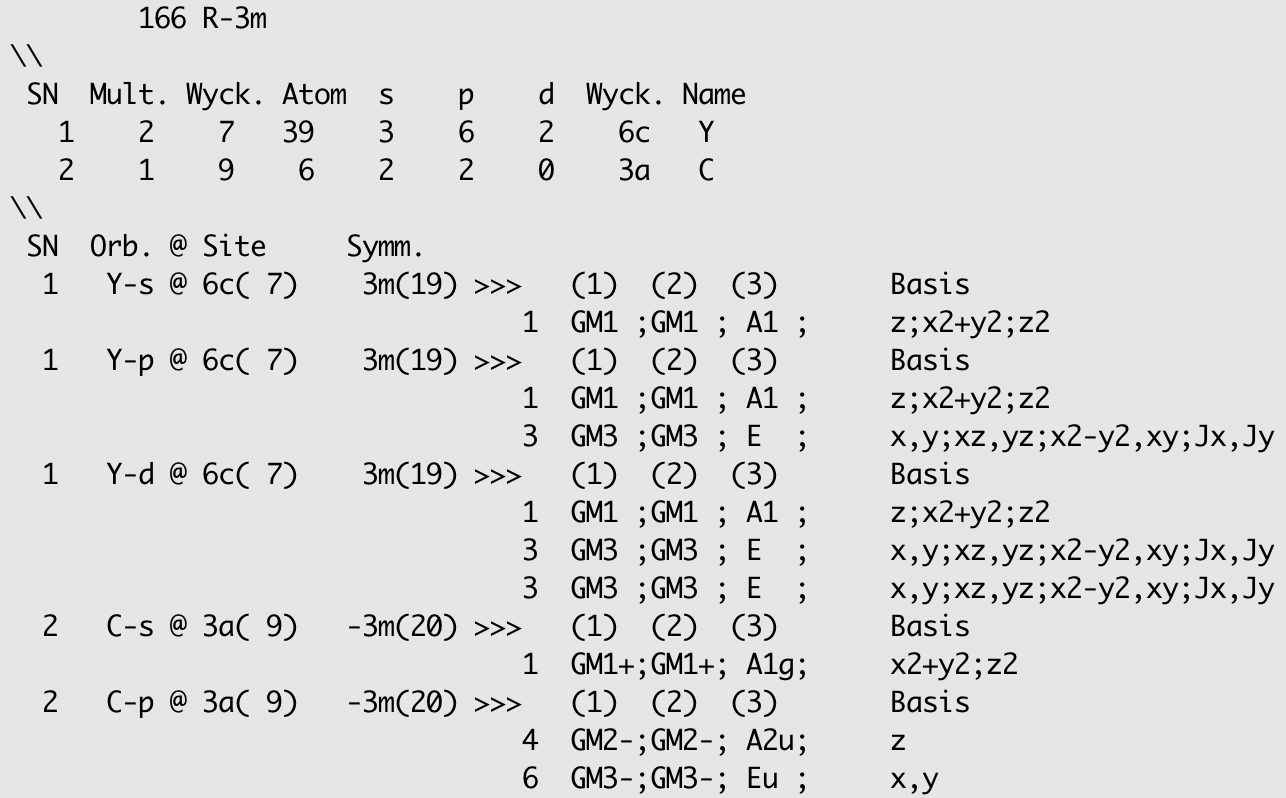}
  \caption{The standard output of \webposbr~for Y$_2$C. The notations of point group irreps under ``(1)'' are used in Bilbao Crystallographic Server~\cite{aroyo2011crystallography,aroyo2006bilbao1,aroyo2006bilbao2}. The notations under ``(2)'' and ``(3)'' follow Ref.~\cite{intsch}}
  \label{fig:Y2C}
\end{figure}

\subsection{Irrep and BR analysis of Ca$_2$N and LaCl}
\label{sup:b}

Ca$_2$N and LaCl both crystallize in a hexagonal layered structure of space group $R\bar3m$ (\#166) with only three atoms (two calcium and one nitrogen atoms) and four atoms (two lanthanum and two chlorine atoms) 
in the primitive rhombohedral unit cells, respectively. 
The basic building blocks of Ca$_2$N and LaCl are tightly bound triple-layer and quadruple-layer structures, respectively.
For the triple-layer structure in Ca$_2$N, the nitrogen layer is sandwiched between two calcium layers in the sequence of Ca-N-Ca, while two hexagonal lanthanum layers are sandwiched between two hexagonal chlorine layers in the sequence of Cl-La-La-Cl for the quadruple-layer structures.
The adjacent triple layers and quadruple layers are stacked loosely, resulting in the floating electrons moving into the interlayer gaps [Fig. 2(d)] and intralayer gaps [Fig. 2(e)], respectively. Atomic positions and aBRs of Ca$_2$N and LaCl are shown in Table~\ref{table:s1} and Table~\ref{table:s3}, respectively. 

The band structure of Ca$_2$N shows three valence bands dominated by N $p$ orbitals in the energy range -4 eV to -1.5 eV, and a dispersive band traversing the Fermi level, as shown in Fig. 2(a). 
To understand the \ele~essence in terms of TQC theory, we calculate the irreps and BRs for the four bands near the Fermi level, as shown in Table~\ref{table:s2}. It is clear that the BR $A_{1g}@3b$ from these bands is an eBR, and generated by electrons located at the vacancy $3b$, suggesting that Ca$_2$N  is an \ele.

The band structure of \ll~is shown in Fig. 2(b), and irreps and BRs for the three bands near the Fermi level of~\ll~are shown in Table~\ref{table:s4}. Since there are band inversions at $\Gamma$ and T, the irreps for the highest band at $\Gamma$ and T are unknown.  After carefully analyzing the BRs, one can get that the unknown irreps should be GM1+(1) and T2--(1). Lastly, we find that these bands support an eBR $A_{1g}@3b$ generated by electrons located at the vacancy $3b$, which suggests that~\ll~is an \ele.

\begin{table*}[!htb]
\caption{
Atomic positions and aBRs of Ca$_2$N (space group $R\bar3m$).
}\label{table:s1}
\begin{tabular}{c|c|c|c|cc|c}
\hline
\hline
  Atom &WKS($q$) &  Site symm.&Conf. &\multicolumn{2}{c|}{Irreps($\rho$)}& aBRs($\rho@q$)\\
\hline
Ca&$6c$& $3m$  & $3p^64s^2$ &$s$&$:A_1$&$A_1$@$6c$  \\
\cline{5-7}
 &    & &  &$p_z$&$:A_1$&$A_1$@$6c$  \\
 &    & &  &$p_x,p_y$&$:E$&$E$@$6c$   \\
  \hline
N &$3a$  & $-3m$ & $2s^22p^3$ & $s$&$:A_{1g}$& $A_{1g}$@$3a$\\
\cline{5-7}
        &    &               &           &$p_z$&$:A_{2u}$ &$A_{2u}$@$3a$   \\
        &    &               &           & $p_x,p_y$&$:E_u $&  $E_u$@$3a$  \\
\hline
\hline
\end{tabular}
\end{table*}

\begin{table*}[!h]
\caption{
Irreps and BRs for the four bands near the Fermi level of Ca$_2$N.
}\label{table:s2}
\begin{tabular}{c|c|c|c|c}
\hline
\hline
 & $\Gamma$ & T & F & L  \\
\hline
& GM1+(1)& T2--(1) &F2--(1)& L2--(1)\\
Bands  & GM2--(1)& T2--(1)&F2--(1)& L2--(1) \\
& GM3--(2) & T3--(2) &F1--(1)& L1--(1) \\
&       &       &F1+(1)& L2--(1) \\
\hline
\hline
BRs \\
\hline
$A_{2u}@3a$& GM2--(1)& T2--(1) &F2--(1)& L2--(1)\\
\hline
$E_{u}@3a$ & GM3--(2)& T3--(2)&F1--(1) $\oplus$ F2--(1)& L1--(1)$\oplus$ L2--(1) \\
\hline
${\color{blue}A_{1g}@3b}$ & {\color{blue}GM1+(1)} & {\color{blue}T2--(1)} &{\color{blue}F1+(1)}& {\color{blue}L2--(1)} \\
\hline
\hline
\end{tabular}
\end{table*}

\begin{table*}[t]
\caption{
Atomic positions and aBRs of LaCl (space group $R\bar3m$). 
}\label{table:s3}
\begin{tabular}{c|c|c|c|cc|c}
\hline
\hline
Atom &WKS($q$) &  Site symm.& Conf. &\multicolumn{2}{c|}{Irreps($\rho$)}& aBRs($\rho@q$)\\
\hline
La& $6c$& $3m$  &$5s^25p^66s^25d^1$ &$s$&$: A_1$  &$A_1$@$6c$\\
\cline{5-7}
    &        &  &      &$p_z$&$: A_1$                           &$A_1$@$6c$\\
    &        &  &      &$p_x,p_y$&$: E$&$E$@$6c$\\
\cline{5-7}    
    &        &  &      &$d_{z^2}$&$: A_1$                           &$A_1$@$6c$\\
    &        &  &      &$d_{xy},d_{xz},d_{yz},d_{x^2-y^2}$&$: E$&$E$@$6c$\\
  \hline
Cl&$6c$& $3m$  & $3s^23p^5$ &$s$&$: A_{1}$                    &$A_1$@$6c$\\
\cline{5-7}
    &        &  &        &$p_z$&$: A_{1}$                &$A_1$@$6c$\\
    &        &  &        &$p_x, p_y$&$: E$                &$E$@$6c$\\
\hline
\hline
\end{tabular}
\end{table*}

\begin{table*}[h]
\caption{
Irreps and BRs for the three bands near the Fermi level of LaCl. The question mark (``?'') stands for any one-dimensional (1D) irrep (\ie the unknown irrep) at $\Gamma$ and T. After analyzing the BRs, one then can conjecture that the unknown irrep should be GM1+(1) and T2--(1).
}\label{table:s4}
\begin{tabular}{c|c|c|c|c}
\hline
\hline
 & $\Gamma$ & T & F &L    \\
\hline
Bands& GM1+(1)&T2--(1)&   F1+(1)&  L2--(1)\\
     & GM2--(1)&T1+(1)&   F1+(1)&  L1+(1)\\
     &{\color{red} ? (1)}&{\color{red} ? (1)}&   F2--(1)&  L2--(1)\\
\hline
\hline
BRs (one of cases)                     \\
\hline
$A_{1g}@3a$ & GM1+(1) & T1+(1) &F1+(1)&L1+(1) \\
\hline
$A_{2u}@3a$ & GM2--(1) & T2--(1) &F2--(1)& L2--(1) \\
\hline
${\color{blue}A_{1g}@3b}$ & {\color{blue}GM1+(1)} & {\color{blue}T2--(1)} &{\color{blue}F1+(1)}&{\color{blue}L2--(1)} \\
\hline
\hline
\end{tabular}
\end{table*}

\subsection{Irrep and BR analysis of C12A7}
\label{sup:c}

C12A7 with low work function (about 0.6 eV) is a room-temperature \ele, whose unusual properties are due to its special crystalline structure with the space group of $I\bar43d$ (\#220). The stoichiometric cubic unit cell is represented by the formula [Ca$_{24}$Al$_{28}$O$_{64}$]$^{4+}\cdot$ [O$^{2-}]_2$. The first term [Ca$_{24}$Al$_{28}$O$_{64}$]$^{4+}$ denotes a positively charged framework built of twelve cages, while the second term [O$^{2-}]_2$ represents two extra-framework oxygen ions occupying two of the twelve cages. By reduction treatment, the extra-framework oxygen ions can be replaced with electrons acting as anions, leading to the formation of the \ele~C12A7. Atomic positions and aBRs of C12A7 are shown in Table~\ref{table:s5}. 

The band structure of C12A7 is shown in Fig. 2(c), irreps and BRs for the six bands near the Fermi level of C12A7 are shown in Table~\ref{table:s6}. This six bands support an eBR $A_{1}@12b$ located at the vacancy $12b$, as shown in Fig. 2(f), which confirms that C12A7 is an \ele.

\begin{table*}[!htb]
\caption{
Atomic positions and aBRs of C12A7 (space group $I\bar43d $).
}\label{table:s5}
\begin{tabular}{c|c|c|c|cc|c}
\hline
\hline
Atom &WKS($q$) &  Site symm.& Conf. &\multicolumn{2}{c|}{Irreps($\rho$)}& aBRs($\rho@q$)\\
\hline
Ca&$24d$& $2$  & $3p^64s^2$ &$s$&$:A$&$A$@$24d$  \\
\cline{5-7}
&   & &   &$p_z$&$:A $&$A$@$24d$ \\
&   & &   &$p_x,p_y$&$:B $&$B$@$24d$   \\
  \hline
Al(1) &$12a$  & $-4$ & $3s^23p^1$ & $s$&$:A$& $A$@$12a$\\
\cline{5-7}
        &    &               &           &$p_z$&$:B$ &$B$@$12a$   \\
        &    &               &           & $p_x$&$:\prescript{2}{}E$&  $^2E$@$12a$  \\
        &    &               &           & $p_y$&$: \prescript{1}{}E$&  $^1E$@$12a$  \\
   \hline
Al(2) &$16c$  & $  3$ & $3s^23p^1$ & $s$&$:A_1$& $A_1$@$16c$\\
\cline{5-7}
        &    &               &           &$p_z$&$:A_1$ &$A_1$@$16c$   \\
        &    &               &           &$p_y$&$:\prescript{2}{}E$ &$^2E$@$16c$   \\
        &    &               &           & $p_x$&$:\prescript{1}{}E$&  $^1E$@$16c$  \\
 \hline
O(1) & $16c$  & $  3$ &$2s^22p^4$ & $s$&$:A_1$& $A_1$@$16c$\\
\cline{5-7}
        &    &               &           &$p_z$&$:A_1$ &$A_1$@$16c$   \\
        &    &               &           &$p_y$&$:\prescript{2}{}E $ &$^2E$@$16c$   \\
        &    &               &           & $p_x$&$:\prescript{1}{}E$&  $^1E$@$16c$  \\
  \hline
O(2) &$48e$  & $1$ & $2s^22p^4$ & $s$&$:A$& $A$@$48e$\\
\cline{5-7}
        &    &               &           &$p_z,p_x,p_y$&$:A$&$A$@$48e$ \\
\hline
\hline
\end{tabular}
\end{table*}

\begin{table*}[!h]
\caption{
Irreps and BRs for the six bands near the Fermi level of C12A7.
}\label{table:s6}
\begin{tabular}{c|c|c|c|c}
\hline
\hline
 & $\Gamma$ & H & P & N  \\
\hline
              &GM1(1) & H4H5(6) &P3(4)&  N1(2)\\
Bands    &GM5(3)&                  & P2(2) & N1(2)\\
             & GM3(2) &                   &          & N1(2) \\
\hline
\hline
BRs \\
\hline
${\color{blue}A_{1}@12b}$ & {\color{blue}GM1(1)$\oplus$ GM5(3)$\oplus$GM3(2)} & {\color{blue}H4H5(6)} &{\color{blue}P3(4)$\oplus$P2(2)}& {\color{blue}3N1(2)} \\
\hline
\hline
\end{tabular}
\end{table*}

\subsection{The evolution of mBJ band structures of $A_2B$}
\label{sup:d}

$A_2B$ ($A=$Ca, Sr, and Ba; $B=$ As, Sb, and Bi) is a large family of isostructural materials hosting similar electronic structures.
They are known as \eles~before, and recently draw intensive research interest due to the presence of nontrivial band topology. In order to better understand the 
relation between these two different properties, we calculate the mBJ band structures of $A_2B$, as shown in Fig. \ref{figs2}. Although increasing the weights of the pnicogen elements doesn't change the band structure too much, the heavier alkaline earth metal in $A_2B$ leads to the larger band inversion. 

\begin{figure*}[!htb]
\centering
\includegraphics[width=15cm]{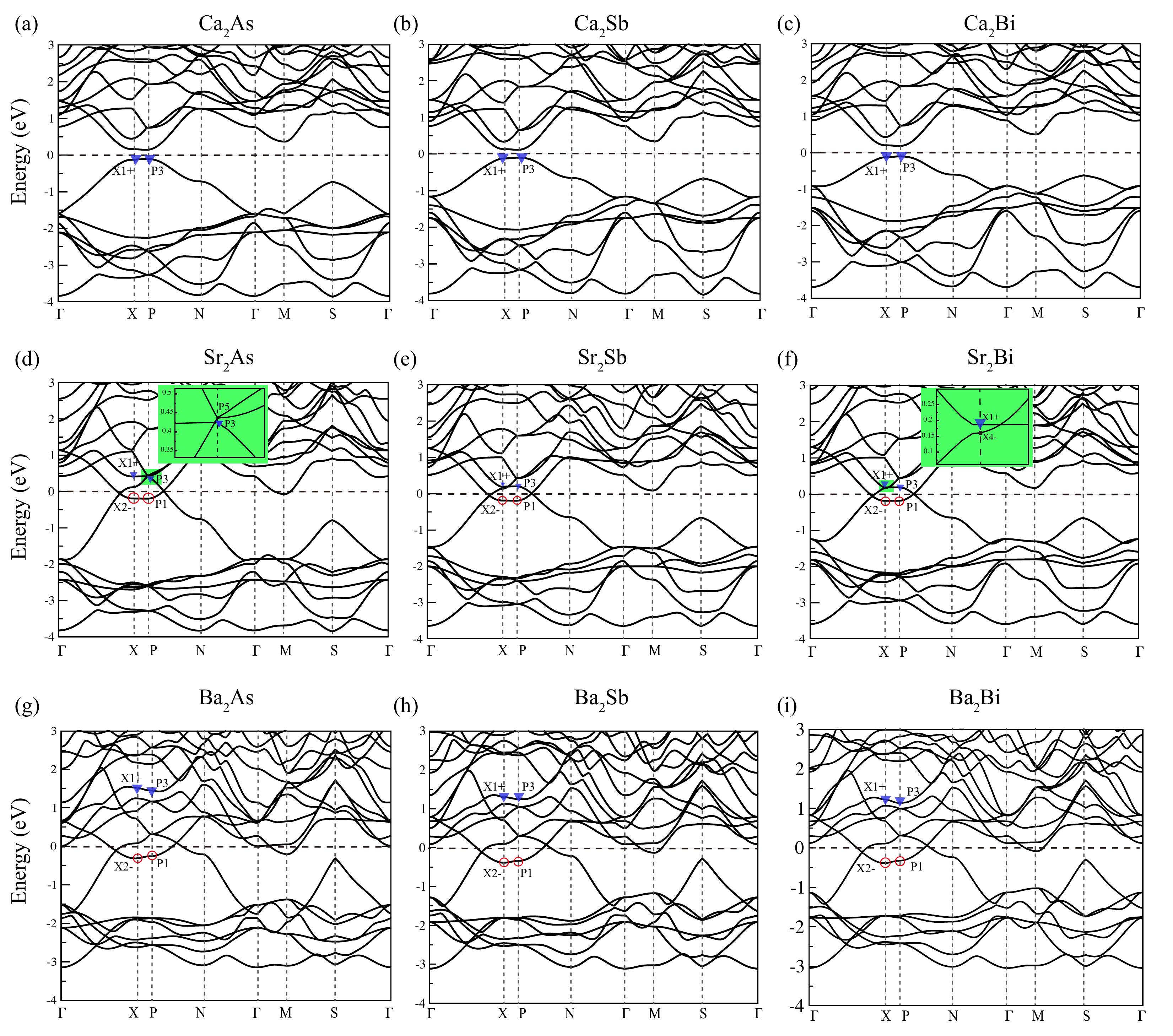}
\caption{(Color online)
The calculated band structures of $A_2B$ family. For the trivial band structures without band inversion, the bands marked by the blue-filled triangles are occupied by charges at vacancies, and the bands marked by red-hollow circles unoccupied. It shows that the band inversion increases as the weights of the pnicogen elements increase.
}
\label{figs2}
\end{figure*}

\subsection{The absorption of hydrogen atoms in Ba$_3$CrN$_3$}
\label{sup:e}

Table \ref{table:s7} shows the aBRs of Ca$_5$P$_3$, which are used to understand the absorption of hydrogen atoms in this material.  
Similar to Ca$_5$P$_3$, we perform the BR analysis on Ba$_3$CrN$_3$ with the space group of P6$_3$/m (\#176), which was previously predicted to be a 1D topological \ele~with anomalous Dirac plasmon~\cite{adp}. 
Fig. \ref{figs3}(a) shows the band structure of Ba$_3$CrN$_3$, where the obtained irreps are marked out. 
Based on the aBRs of Ba$_3$CrN$_3$ shown in Table \ref{table:s8}, the BR analysis shows that the blue-solid bands belong to the eBR of $A_{1g} @ 2b$, induced by excess electrons at the vacancies $2b$. By absorbing hydrogen atoms at the vacancies, these bands are removed from $E_F$ for \bcn H [Fig.~\ref{figs3}(b)]. The insulating band structure is consistent with recent measurements, since the charge-balanced compound~\bcn H has been obtained in experiments~\cite{falb2019ba3crn3h}. So the results of hydrogen absorption clearly show that \eles~exhibit excess electrons at vacancies.

\begin{figure*}[!htb]
\centering
\includegraphics[height=10cm]{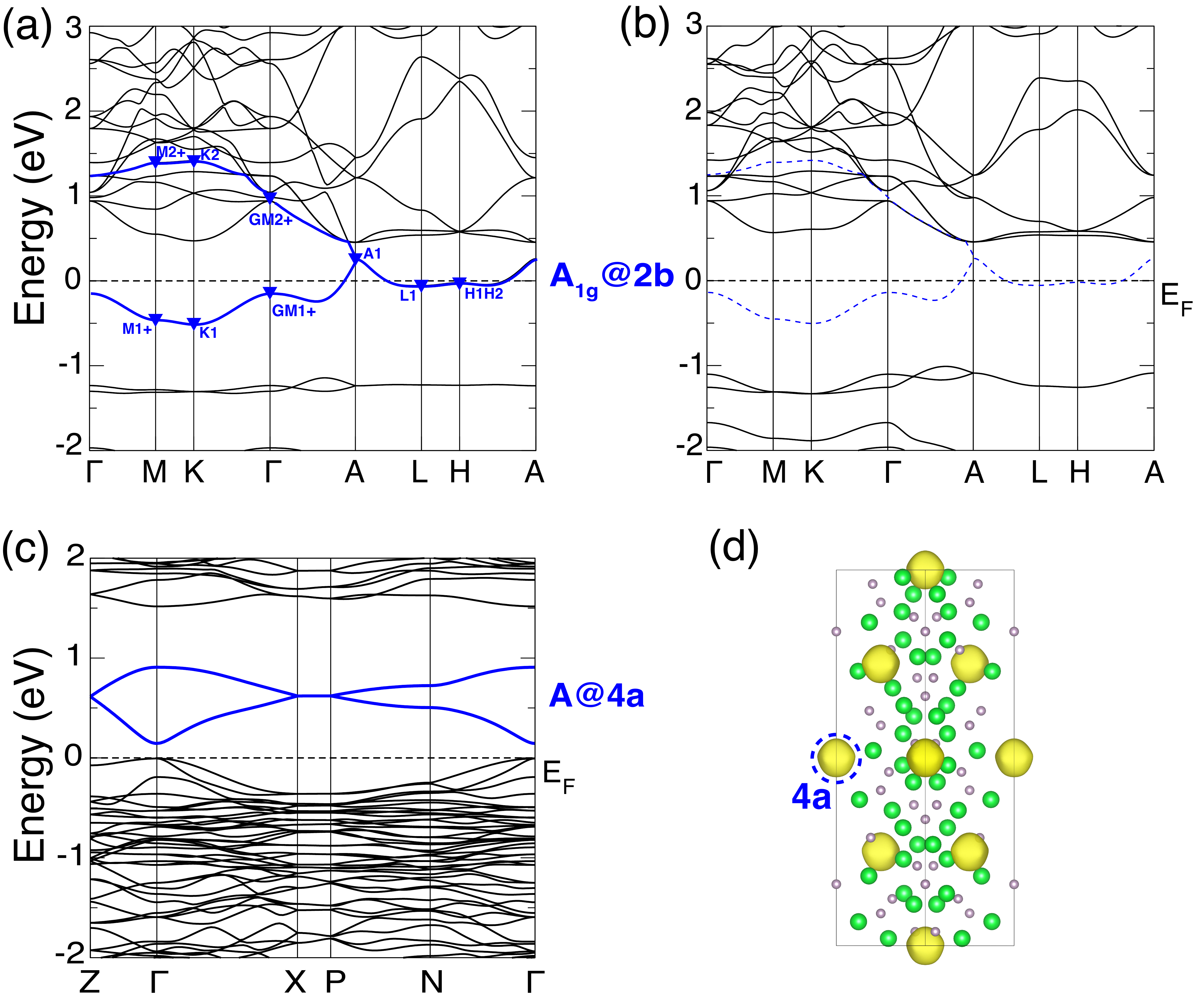}
\caption{(Color online)
The band structures of \bcn~(a), \bcn H~(b) and Sr$_3$P$_2$ (c). The floating bands (blue-solid) denote electronic states of excess electrons, whose eBRs can be well determined by the obtained irreps. To facilitate the comparison, the irreps of particular bands are marked out, and the ``disappeared'' floating bands are depicted by blue-dashed lines in (b) for \bcn H. (d) The PED of Sr$_3$P$_2$.
}
\label{figs3}
\end{figure*}

\begin{table*}[!htb]
\caption{   
Atomic positions and aBRs of Ca$_5$P$_3$ (space group $P6_3/mcm$).
}\label{table:s7}
\begin{tabular}{c|c|c|c|cc|c}
\hline      
\hline      
Atom &WKS($q$) &  Site symm.& Conf. &\multicolumn{2}{c|}{Irreps($\rho$)}& aBRs($\rho@q$)\\
\hline      
Ca(1)&$4d$& $32$   & $3p^64s^2$ &$s$&$: A_1$    &$A_1$@$4d$\\
\cline{5-7}
    &        &  &      &$p_z$&$: A_2$ &$A_2$@$4d$\\
    &        &  &      &$p_x,p_y$&$: E$&$E$@$4d$\\
  \hline    
Ca(2)&$6g$& $mm2$  & $3p^64s^2$&$s$&$: A_1  $    &$A_1$@$6g$\\
\cline{5-7}
    &        &  &     &$p_z$&$: A_1$    &$A_1$@$6g$\\
    &        &  &     &$p_y$&$: B_{2}$ &$B_2$@$6g$\\
    &        &  &     &$p_x$&$: B_{1}$ &$B_1$@$6g$\\
\hline
P&$6g$& $mm2$  & $3s^23p^3$ &$s$&$: A_1  $    &$A_1$@$6g$\\
\cline{5-7}
    &        &  &     &$p_z$&$: A_1$    &$A_1$@$6g$\\
    &        &  &     &$p_y$&$: B_{2}$ &$B_2$@$6g$\\
    &        &  &     &$p_x$&$: B_{1}$ &$B_1$@$6g$\\
\hline      
\hline      
\end{tabular}
\end{table*}

\begin{table*}[!htb]
\caption{   
Atomic positions and aBRs of Ba$_3$CrN$_3$ (space group $P6_3/m$).
}\label{table:s8}
\begin{tabular}{c|c|c|c|cc|c}
\hline      
\hline      
Atom &WKS($q$) &  Site symm.& Conf. &\multicolumn{2}{c|}{Irreps($\rho$)}& aBRs($\rho@q$)\\
\hline      
Ba&$6h$& $m$   & $5s^26s^25p^6$ &$s$&$: A'$    &$A'$@$6h$\\
\cline{5-7}
    &        &  &      &$p_x,p_y$&$: A'$ &$A'$@$6h$\\
   &        &  &      &$p_z$&$: A''$&$A''$@$6h$\\
  \hline    
Cr&$2c$& $\bar6$  & $3d^54s^1$&$s$&$: A'  $    &$A^{'}$@$2c$\\
\cline{5-7}
    &        &  &     &$p_z$&$: A''$    &$A''$@$2c$\\
    &        &  &     &$p_x$&$: \prescript{2}{}E'$ &$ \prescript{2}{}E'$@$2c$\\
    &        &  &     &$p_y$&$: \prescript{1}{}E'$ &$ \prescript{1}{}E'$@$2c$\\
\cline{5-7}
    &        &  &     &$d_{z^2}$&$: A'$    &$A^{'}$@$2c$ \\
    &        &  &     &$d_{x^2-y^2}$&$: \prescript{2}{}E'$ &$ \prescript{2}{}E'$@$2c$\\
    &        &  &     &$d_{xz}$&$: \prescript{2}{}E''$ &$ \prescript{2}{}E''$@$2c$\\
    &        &  &     &$d_{xy}$&$: \prescript{1}{}E'$ &$ \prescript{1}{}E'$@$2c$\\
    &        &  &     &$d_{yz}$&$: \prescript{1}{}E''$ &$ \prescript{1}{}E''$@$2c$\\
\hline
N&$6h$& $m$  & $2s^22p^3$ &$s$&$: A'  $    &$A'$@$6h$\\
\cline{5-7}
    &        &  &     &$p_x,p_y$&$: A'$    &$A'$@$6h$\\
    &        &  &     &$p_z$&$: A''$ &$A''$@$6h$\\
\hline      
\hline      
\end{tabular}
\end{table*}

\subsection{The BR analysis of the predicted \eles}
\label{sup:f}

In order to further understand the eBRs contributed from the states at vacancies, we tabulate the atomic positions and aBRs for NaBaO, K$_2$O, Sr$_6$GaN$_5$ and Sr$_3$P$_2$ in Tables~\ref{table:s9},~\ref{table:s10}, ~\ref{table:s11} and~\ref{table:s12}, respectively. We here focus on another predicted \ele, \ie~Sr$_3$P$_2$. The generalized gradient approximation band structure of Sr$_3$P$_2$ with the space group of I$\bar{4}$2d (\#122) is presented in Fig.~\ref{figs3}(c). The BR analysis indicates that two blue-colored bands (above $E_F$) form the eBR of $A @ 4b$. The PED of the two bands is plotted in Fig.~\ref{figs3}(d), agreeing well with the BR analysis. 

Since the conduction bands of both Sr$_3$P$_2$ and Sr$_6$GaN$_5$ are well separated from other bands, and the charge centers of these conduction bands are well defined at the vacancies, we propose that Sr$_3$P$_2$ and Sr$_6$GaN$_5$ are potential \eles; namely, they could be \eles~after electron doping. 
As shown in Figs. 5(a,c) in the main text, the floating bands of K$_2$O and Sr$_6$GaN$_5$ are the highest conduction bands. To occupy these bands, electron doping with suitable elements is employed. For K$_2$O and Sr$_6$GaN$_5$, 50\% oxygen and 50\% gallium are doped with F and Ge, respectively.
The electronic structures of K$_2$O$_{0.5}$F$_{0.5}$ and Sr$_6$Ga$_{0.5}$Ge$_{0.5}$N$_5$ are studied with the virtual crystal approximation method, and the results are shown in Fig. \ref{figs4}. It is clearly that the floating bands are half filled, and the overall band structures almost don't change except an overall band structure shift.

\begin{table*}[!htb]
\caption{   
Atomic positions and aBRs of NaBaO (space group $P4/nmm$).
}\label{table:s9}
\begin{tabular}{c|c|c|c|cc|c}
\hline      
\hline      
Atom &WKS($q$) &  Site symm.& Conf. &\multicolumn{2}{c|}{Irreps($\rho$)}& aBRs($\rho@q$)\\
\hline      
Ba&$2c$& $4mm$   & $5s^26s^25p^6$ &$s$&$: A_1$    &$A_1$@$2c$\\
\cline{5-7}
    &        &  &      &$p_z$&$: A_1$ &$A_1$@$2c$\\
    &        &  &      &$p_x,p_y$&$: E$&$E$@$2c$\\
  \hline    
O&$2c$& $4mm$  & $2s^22p^4$&$s$&$: A_1 $    &$A_1$@$2c$\\
\cline{5-7}
    &        &  &      &$p_z$&$: A_1$ &$A_1$@$2c$\\
    &        &  &      &$p_x,p_y$&$: E$& $E$@$2c$ \\
\hline
Na&$2c$& $4mm$  & $3s^1$ &$s$&$: A_1$ &$A_1$@$2c$\\
\hline      
\hline      
\end{tabular}
\end{table*}

\begin{table*}[!htb]
\caption{   
Atomic positions and aBRs of K$_2$O (space group $Fm\bar3m$).
}\label{table:s10}
\begin{tabular}{c|c|c|c|cc|c}
\hline      
\hline      
Atom &WKS($q$) &  Site symm.& Conf. &\multicolumn{2}{c|}{Irreps($\rho$)}& aBRs($\rho@q$)\\
\hline      	
O&$4a$& $m\bar3m$   & $3p^64s^2$ &$s$&$: A_{1g}$    &$A_{1g}$@$4a$\\
\cline{5-7}
    &        &  &      &$p_x,p_y,p_z$&$: T_{1u}$&$T_{1u}$@$4a$\\
  \hline    
K&$8c$& $\bar43m$  & $3p^64s^2$&$s$&$: A_1  $    &$A_{1}$@$8c$\\
\cline{5-7}
    &        &  &     &$p_x,p_y,p_z$&$: T_{1}$   &$T_{1}$@$8c$\\
\hline      
\hline      
\end{tabular}
\end{table*}


\begin{table*}[!htb]
\caption{   
Atomic positions and aBRs of Sr$_6$GaN$_5$ (space group $P6_3/mcm$).
}\label{table:s11}
\begin{tabular}{c|c|c|c|cc|c}
\hline      
\hline      
Atom &WKS($q$) &  Site symm.& Conf. &\multicolumn{2}{c|}{Irreps($\rho$)}& aBRs($\rho@q$)\\
\hline      
Sr&$12k$& $m$   & $5s^26s^25p^6$ &$s$&$: A'$    &$A'$@$12k$ \\
\cline{5-7}
    &        &  &      &$p_x/p_y$&$: A'$ &$A'$@$12k$ \\
  \cline{7-7}
    &        &  &      &$p_z$&$: A''$&$A''$@$12k$ \\
 \hline    
Ga& $2a$& $\bar62m$  &$4s^24p^1$&$s$&$: A^{'}_1  $    &$A^{'}_1$@$2a$\\
\cline{5-7}
    &        &  &     &$p_z$&$: A^{''}_2$    &$ A^{''}_2$@$2a$\\
    &        &  &     &$p_y$&$:E^{'}$ &$E^{'}$@$2a$\\
\cline{5-7} 
    &        &  &     &$d_{z^2}$&$: A^{'}_1 $    &$A^{'}_1$ @$2a$\\
    &        &  &     &$d_{x^2-y^2},d_{xy}$&$:E^{'}$ &$E^{'}$@$2a$\\
    &        &  &     &$d_{xz},d_{yz}$&$: E^{''}$ &$E^{''}$@$2a$\\
\hline
N(1)&$4d$& $32$  & $2s^22p^3$&$s$&$: A_1  $    &$A_1$@$4d$\\
\cline{5-7}
    &        &  &     &$p_z$&$: A_2$    &$ A_2$@$4d$\\
    &        &  &     &$p_x,p_y$&$:E$ &$E$@$4d$\\
\hline
N(2)&$6g$& $mm2$  & $2s^22p^3$ &$s$&$: A_1  $    &$A_1  $@$6g$\\
\cline{5-7}
    &        &  &     &$p_z$&$: A_1$    &$A_1  $@$6g$\\
    &        &  &     &$p_y$&$: B_{2}$ &$B_2  $@$6g$\\
    &        &  &     &$p_x$&$: B_{1}$ &$B_1  $@$6g$\\
\hline      
\hline      
\end{tabular}
\end{table*}

\begin{table*}[!htb]
\caption{   
Atomic positions and aBRs of Sr$_3$P$_2$ (space group I$\bar{4}$2d).
}\label{table:s12}
\begin{tabular}{c|c|c|c|cc|c}
\hline      
\hline      
Atom &WKS($q$) &  Site symm.& Conf. &\multicolumn{2}{c|}{Irreps($\rho$)}& aBRs($\rho@q$)\\
\hline      	
Sr(1)&$16e$& $1$   & $5s^26s^25p^6$ &$s/p_x/p_y/p_z$ &$: A$    &$A$@$16e$ \\
\hline      	                     
Sr(2)&$16e$& $1$   & $5s^26s^25p^6$ &$s/p_x/p_y/p_z$ &$: A$    &$A$@$16e$ \\
\hline      	                     
Sr(3)&$16e$& $1$   & $5s^26s^25p^6$ &$s/p_x/p_y/p_z$ &$: A$    &$A$@$16e$ \\
\hline      
P(1)&$8c$& $2$  & $3s^23p^3$ &$s/p_z$&$: A$    &$A$@$8c$\\
\cline{5-7}
  &    &      &            &$p_x/p_y$&$: B$    &$B$@$8c$\\
\hline      
P(2)&$8d$& $2$  & $3s^23p^3$ &$s/p_z$&$: A$    &$A$@$8d$\\
\cline{5-7}
  &    &      &            &$p_x/p_y$&$: B$    &$B$@$8d$\\
\hline      
P(3)&$16e$& $1$  & $3s^23p^3$ &$s/p_x/p_y/p_z$ &$: A$    &$A$@$16e$\\
\hline      
\hline      
\end{tabular}
\end{table*}

\begin{figure*}[!htb]
\centering
\includegraphics[height=7cm]{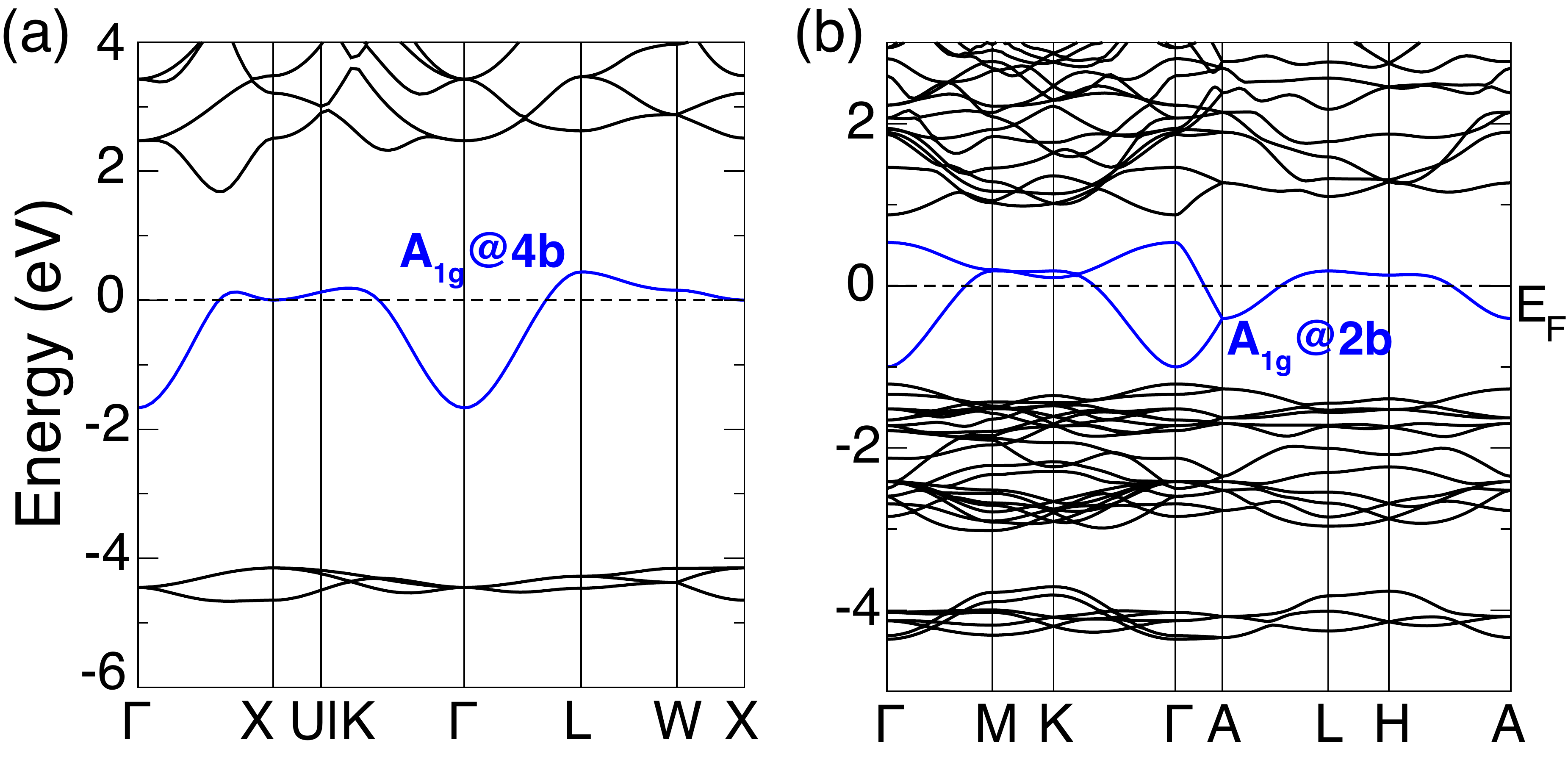}
\caption{(Color online)
(a,b) The band structures of K$_2$O$_{0.5}$F$_{0.5}$ and Sr$_6$Ga$_{0.5}$Ge$_{0.5}$N$_5$, respectively.
}
\label{figs4}
\end{figure*}

\clearpage

\end{widetext}
\end{document}